\documentclass[sigconf]{acmart}
\usepackage{multirow}   
\usepackage{enumitem}
\usepackage{tabularx}
\usepackage{graphicx}
\usepackage{caption}
\usepackage{placeins}
\usepackage{stfloats}
\usepackage{booktabs}    
\usepackage{array}       
\usepackage{threeparttable} 

\AtBeginDocument{%
  }

\setcopyright{acmlicensed}
\copyrightyear{2027}
\acmYear{2027}
\acmDOI{}

\acmISBN{}

\begin{document}

\title{Aligning Performance with Contribution: Towards Contribution-Aware Fair Recommendation}


\author{Shuai Zhang}
\affiliation{%
  \institution{Shanghai University of Finance and Economics}
  \country{China}
}
\email{2023214701@stu.sufe.edu.cn}

\author{Hui Fang}
\affiliation{%
  \institution{Shanghai University of Finance and Economics}
  \country{China}
}
\email{fang.hui@mail.shufe.edu.cn}

\author{Zhu Sun}
\affiliation{%
  \institution{Singapore University of Technology and Design}
  \country{Singapore}
}
\email{zhu_sun@sutd.edu.sg}

\renewcommand{\shortauthors}{}

\begin{abstract}
Existing research on user fairness in recommender systems has developed diverse objectives. However, it has paid limited attention to a distinct distributive perspective: whether users' contributions to model learning should be reflected in the recommendation benefits they receive. We argue that, in addition to existing fairness protections, a fair system may account for the alignment between users' estimated contributions and the recommendation performance they receive. Such alignment can incentivize sustained and informative engagement, thereby supporting a sustainable recommendation ecosystem. To this end, we propose \textbf{Contribution-Performance Fairness}, a novel fairness perspective which requires recommendation performance to be aligned with estimated contribution across user groups and to remain equitable among users with comparable contributions within a same group. To instantiate this perspective, we introduce the \textbf{Contribution-Performance Fair Recommender (CPFR)}, a framework applicable to different backbone recommenders. CPFR constructs ordered user groups from a training-dependent contribution considering interaction volume, loss alignment, and optimization intensity, and jointly optimizes recommendation accuracy with the two fairness requirements. A game-theoretic analysis shows that such alignment can strengthen contribution incentives and improve system-level recommendation accuracy under voluntary contribution. Experiments on three datasets and three backbone models demonstrate that CPFR achieves a strong accuracy--fairness trade-off under the proposed operational metric. 
\end{abstract}

\ccsdesc[500]{Information systems~Recommender systems}
\keywords{Fair Recommendation, Recommender Systems, User Contribution}

\maketitle

\section{INTRODUCTION}
\label{sec:introduction}

Recommender systems (RSs) have become the cornerstone of modern information services, alleviating information overload and driving user engagement across e-commerce, social media, and streaming platforms~\cite{ricci2015recommender}. As these systems increasingly mediate access to opportunities and content, ensuring algorithmic fairness has emerged as a critical research frontier~\cite{wang2023survey,deldjoo2023fairness}. Many fairness-aware recommendation methods seek parity in outcomes across groups defined by sensitive or behavioral attributes~\cite{barocas2019fairness,singh2018fairness,pmlr-v81-ekstrand18b}. Complementary work considers non-uniform fairness based on target distributions, relevance, or user similarity~\cite{deldjoo2021flexible,biega2018equity,rampisela2026measuring}. Together, these studies provide diverse definitions and perspectives on fairness in RSs.

However, existing work pays limited attention to a fundamental asymmetry: users are not merely recipients but active contributors of data. High-contributing users provide abundant, high-quality interactions that are critical for model training, whereas others may contribute sparse or noisy signals. Existing methods do not directly address whether a user's contribution to model learning should be reflected in the recommendation performance the user receives. When contribution heterogeneity is ignored, parity-oriented treatments may create a structural misalignment by suppressing the performance of high-contribution users to match lower-contribution users, thereby weakening incentives and potentially undermining long-term ecosystem health.

We advocate explicitly incorporating estimated, training-dependent user contribution into model optimization to achieve \textit{contribution-aligned benefit}. We introduce \textbf{Contribution-Performance Fairness}, a novel fairness perspective in which users with comparable estimated contributions should receive comparable performance, and higher-contribution groups should receive no worse average performance compared with lower-contribution ones. Unlike immutable demographic attributes (e.g., race or gender), estimated user contribution depends on engagement and may change over time. Users can therefore receive better recommendation performance by contributing more informative data. Notably, this principle is intended for settings with voluntary, model-relevant feedback, and complements rather than replaces equality-based or protected-group fairness. Low contribution should not reduce service below a reasonable quality floor, since sparse activity may reflect cold-start status, limited access, or privacy choices rather than unwillingness to contribute.


Realizing this perspective introduces two key challenges:

\textbf{(C1) Estimating a Multi-dimensional and Ordered Contribution Structure.} User contribution is not directly observed and may be reflected by multiple signals from user data and model training. Deriving a stable contribution estimate and ordered groups from these signals is non-trivial. \textbf{(C2) Multi-objective Optimization for Discrete Groups.} Enforcing contribution-aligned fairness requires jointly optimizing recommendation accuracy, cross-group alignment, and within-group equity, which introduces non-trivial optimization trade-offs and potential gradient conflicts.



To address the above challenges, we propose \textbf{CPFR} (\textbf{C}ontribution-\textbf{P}erformance \textbf{F}air \textbf{R}ecommender), a model-agnostic framework that operationalizes contribution-based fairness through a unified learning paradigm for contribution-performance-aligned recommendation. To address C1, CPFR uses multi-dimensional signals derived from each user's data and model-training dynamics as proxies for user contribution. It then partitions users into ordered contribution-aware groups for subsequent optimization. To address C2, CPFR incorporates two complementary fairness objectives directly into model training. First, \textbf{cross-group alignment} enforces a monotonic relationship between contribution and recommendation performance, encouraging that higher-contribution groups achieve no worse outcomes compared with the lower-contribuiton ones. Second, \textbf{within-group equity} minimizes performance disparities among users with similar contribution levels. These objectives are instantiated through two dedicated regularization terms, which are jointly optimized with the recommendation loss in a multi-objective framework. By integrating cross-group ordering constraints and within-group variance reduction into the gradient updates, CPFR balances fairness and accuracy without ad hoc post-processing.

We evaluate CPFR on three benchmark datasets and three backbone models using accuracy and fairness metrics as well as comprehensive recommendation quality ($Q$) as a holistic measure of the accuracy–fairness balance. CPFR improves $Q$ over fairness baselines in most settings while maintaining competitive recommendation accuracy. Under the idealized allocation model in Sec.~\ref{sec:theory}, our theoretical analysis shows that contribution-proportional allocation yields higher equilibrium contribution and system accuracy than egalitarian allocation.

In summary, our contributions are:
\vspace{-3pt}
\begin{itemize}
    \item We introduce Contribution-Performance Fairness, a novel fairness perspective requiring users with comparable estimated contributions to receive comparable recommendation performance, while ensuring that groups with higher average contributions receive no worse average performance than groups with lower average contributions.
    \item We propose CPFR, which estimates user contribution from interaction volume and training dynamics, partitions users into ordered groups, and jointly optimizes differentiable objectives for cross-group alignment and within-group equity.
    \item We conduct a theoretical game-theoretic analysis under an idealized allocation model, showing that contribution-proportional allocation induces higher equilibrium user contributions and, consequently, higher aggregate recommendation performance than equal-performance allocation.
    \item Experiments on three datasets and three backbone models show that CPFR improves Comprehensive Recommendation Quality ($Q$) over fairness baselines in most settings while maintaining competitive accuracy.
\end{itemize}





\vspace{-12pt}
\section{RELATED WORK}
\textbf{User Fairness in Recommender Systems.}
Fairness in recommender systems has attracted significant research attention in recent years~\cite{wang2023survey,deldjoo2023fairness}. Existing approaches can be broadly categorized into group fairness and individual fairness. Traditional group fairness methods aim to ensure equitable treatment across user groups defined by sensitive attributes such as gender, age, or race~\cite{barocas2019fairness,ntoutsi2020bias}. Common formulations include demographic parity and equalized odds~\cite{hardt2016equality,singh2018fairness,pmlr-v81-ekstrand18b}, which enforce statistical equality in recommendation outcomes across groups. Following this line, Leonhardt et al.~\cite{leonhardt2018user} first explicitly formulated user fairness concerns, Rahmani et al.~\cite{rahmani2022generalizability} conducted systematic experiments on generalizability, and Bisht et al.~\cite{bisht2025cugf} proposed CUGF for credibility and fairness; additionally, Xuan et al.~\cite{fairns2025} leverages group-specific negative sampling to enhance inter-group fairness. Individual fairness, following Dwork et al.~\cite{dwork2012fairness}, requires similar individuals to receive similar treatment~\cite{TFROM,li2021personalized,wang2023survey}. Recent work also studies the relation between group and individual fairness~\cite{Rampisela_2025,SCRUFD2025}. Non-uniform notions exist as well: Deldjoo et al.~\cite{deldjoo2021flexible} provide an evaluation framework with configurable target distributions; Biega et al.~\cite{biega2018equity} require item attention to be proportional to item relevance over repeated rankings; and Rampisela et al.~\cite{rampisela2026measuring} evaluate user fairness using interaction-history similarity and effectiveness disparity. Our CPFR differs in its fairness basis and target: it estimates each user's contribution to model learning and constrains the recommendation performance received by users across contribution groups.

\textbf{Multi-Objective Optimization in Fair Recommendation.}
Multi-objective optimization is widely adopted to balance accuracy against group fairness or multi-stakeholder interests~\cite{abdollahpouri2020multistakeholder, wu2022multi, mlfrmoip}. Recent works have expanded objectives to include serendipity and user satisfaction~\cite{fasmoea2025, zaizi2025multobj}, or explored the accuracy-fairness trade-off frontier~\cite{lacava2023fomo, liu2022accuracy, chowdhry2024discovering}. Techniques like MGDA~\cite{sener2018multi}, PCGrad~\cite{yu2020gradient}, and CAGrad~\cite{liu2021conflict} are commonly used to mitigate gradient conflicts. Our work operationalizes Contribution-Performance Fairness through a specific multi-objective formulation. Distinct from conventional fairness setups, our optimization simultaneously balances recommendation accuracy, within-group equity, and cross-group alignment. We leverage MoCo~\cite{moco2023} to navigate this optimization landscape, aligning recommendation performance with the estimated user contribution.

\textbf{Contribution-based Incentive Mechanisms.}
The concept of valuing data contributions has been explored in machine learning contexts. Ghorbani and Zou~\cite{ghorbani2019data} proposed Data Shapley for principled data valuation, and Koh and Liang~\cite{koh2017understanding} introduced influence functions to estimate training sample impact. Federated learning faces related challenges in contribution evaluation and fair incentive distribution~\cite{kairouz2021advances}. Shi et al.~\cite{shi2023fairness} surveyed fairness-aware federated learning, Liu et al.~\cite{liu2022gtg} proposed GTG-Shapley for efficient contribution evaluation, and Shi et al.~\cite{shi2022fedfaim} introduced FedFAIM for fair incentives. While federated learning addresses contribution fairness at the client level, our work tackles this at the user level within a single recommender system.

\textbf{Our Work.}
CPFR complements existing non-uniform fairness notions by using a training-dependent contribution estimate as the basis for user-side performance constraints. It enforces comparable performance within contribution groups and monotonic average performance across groups.

\vspace{5pt}


\section{PRELIMINARIES}
\vspace{-2pt}
\subsection{Basic Notations and Definitions}
\vspace{-2pt}

Let $\mathcal{U}$ and $\mathcal{I}$ denote the user and item sets. Users $u_x,u_y\in\mathcal{U}$ are indexed by $x,y$, and $i\in\mathcal{I}$ denotes an item. The interaction matrix $\mathbf{O}\in\{0,1\}^{|\mathcal{U}|\times|\mathcal{I}|}$ is split into training ($\mathbf{O}_{\text{train}}$) and test ($\mathbf{O}_{\text{test}}$) sets. For user $u_x$, $\mathcal{I}_{u_x}=\{i\in\mathcal{I}\mid o_{u_x,i}=1\}$ is the interacted-item set. Let $M$ denote a recommendation metric (e.g., NDCG@$K$ or HR@$K$), $M_{u_x}$ the test performance of $u_x$, $\Theta$ the recommender parameters, and $\ell(u_x,i;\Theta)$ the per-interaction BCE loss.

We partition users into $J$ disjoint groups $\mathcal{G} = \{G_j\}_{j=1}^{J}$ based on their estimated training-contribution scores $C_{u_x}$ (defined in Sec.~3.2), where $G_j \subseteq \mathcal{U}$ and $n_j=|G_j|$. The group contribution $C_{G_j}$ and group performance $M_{G_j}$ are defined as:
\vspace{-5pt}
\begin{equation}
\label{eq:CP}
C_{G_j} = \frac{1}{n_j} \sum_{u_x \in G_j} C_{u_x}, \quad 
M_{G_j} = \frac{1}{n_j} \sum_{u_x \in G_j} M_{u_x}.
\end{equation}

\vspace{-16pt}
\subsection{User Contribution Modeling}
We operationalize a user's contribution to the learning process of a selected recommender through a model-conditioned proxy $C_{u_x}$, integrating three signals derived from its training dynamics:

\textbf{(i) Interaction Volume.} The interaction count is $\mathrm{Vol}_{u_x} = |\mathcal{I}_{u_x}|$.

\textbf{(ii) Loss Alignment.} Following noise-label studies~\cite{ICML2019Noise}, we use negative mean training loss as a proxy for the alignment of a user's interactions with the current model: $\mathrm{Align}_{u_x} = -\frac{1}{|\mathcal{I}_{u_x}|} \sum_{i \in \mathcal{I}_{u_x}} \ell(u_x, i;\Theta)$.

\textbf{(iii) Optimization Intensity.} Inspired by~\cite{Paul2021DeepLO}, we measure local optimization intensity via the average per-sample gradient norm: $\mathrm{Grad}_{u_x} = \frac{1}{|\mathcal{I}_{u_x}|} \sum_{i \in \mathcal{I}_{u_x}} \| \nabla_\Theta \ell(u_x, i; \Theta) \|_2$.

Consistent with prior work~\cite{Paul2021DeepLO}, we collect $\mathrm{Align}_{u_x}$ and $\mathrm{Grad}_{u_x}$ during the early stages of standard training (without fairness constraints) and average results over multiple runs to ensure stability. 

Since these components lie in different numerical scales, we apply z-score normalization across all users. The final \textbf{Individual Contribution Degree} $C_{u_x}$ is a convex combination of these standardized features:
\begin{equation}
\label{eq:contribution}
C_{u_x} = w_1 \overline{\mathrm{Vol}}_{u_x} + w_2 \overline{\mathrm{Align}}_{u_x} + w_3 \overline{\mathrm{Grad}}_{u_x},
\end{equation}
where the overline denotes z-score normalization, and we set $w_1=w_2=w_3=\frac{1}{3}$ by default to represent no preference among the three components. In deployment, weights may be chosen according to the platform's contribution policy. 

Importantly, $C_{u_x}$ is not a measure of a user's intrinsic value or entitlement. It is a model-dependent operational proxy for the estimated contribution of that user's observed interactions to training. CPFR evaluates whether recommendation benefits align with estimated contribution \emph{within a specified model and task setting}, rather than claiming a model-independent ordering of users' contributions. Moreover, CPFR uses $C_{u_x}$ to form ordered contribution strata. Its fairness criteria are therefore based primarily on the induced ordering and grouping, rather than on interpreting absolute score differences as cardinal quantities.

\subsection{Contribution-Performance Fairness}
\label{subsec:contribution-fairness}

We define contribution-performance fairness through two criteria over groups $\mathcal{G}$ formed from estimated contribution levels $C_{u_x}$:
\begin{enumerate}
    \item \textbf{Cross-Group Alignment (Monotonicity).} 
    The system should ensure a monotonic relationship between a group's mean contribution and the average performance its members receive. Formally, for any two groups $G_j, G_k \in \mathcal{G}$, if group $G_j$ has a lower mean contribution than group $G_k$, then the average performance of $G_j$ should not exceed that of $G_k$: 
    \begin{equation}
    \label{eq:alignment}
    C_{G_j} < C_{G_k} \Rightarrow M_{G_j} \leq M_{G_k}.
    \end{equation}

    \item \textbf{Within-group Equity (Homogeneity).} 
    For users within the same contribution group, the system should minimize the disparity in their individual performance. Formally:
    \begin{equation}
    \label{eq:equity}
    \forall G_j \in \mathcal{G}: \text{Var}(\{M_{u_x} \mid u_x \in G_j\}) \text{ is minimized}.
    \end{equation}
\end{enumerate}

To quantify the deviation from these ideals, we define two complementary metrics based on a concrete performance measure $M$:
\begin{itemize}
    \item \textbf{User Group Fairness (UGF).} 
    It measures cross-group rank alignment between contribution and performance. We compute the Spearman correlation~\cite{Spearman2015ThePA} between the rankings of $\{C_{G_j}\}_{j=1}^{J}$ and $\{M_{G_j}\}_{j=1}^{J}$, and normalize it to $[0,1]$:
    \begin{equation}
    \label{eq:ugf}
    \mathrm{UGF} = \frac{\rho_{\text{Spearman}}(\{M_{G_j}\}_{j=1}^{J}, \{C_{G_j}\}_{j=1}^{J}) + 1}{2}.
    \end{equation}
    Higher UGF indicates stronger alignment between contribution and recommendation quality.
    
    \item \textbf{User Individual Fairness (UIF).} 
    It measures within-group equity through the performance disparity among users in the same group. We define within-group unfairness as the average squared difference between all unordered pairs of users:
    \vspace{-10pt}
    
    \begin{equation}
    \label{eq:uif}
    \mathrm{UIF}
    = \frac{1}{J}
    \sum_{j=1}^{J}    \frac{\sum_{\substack{u_x,u_y \in G_j, x<y}}(M_{u_x} - M_{u_y})^2}{n_j(n_j-1)},
    \end{equation}
    \vspace{-10pt}
    where $n_j = |G_j|$. Lower UIF indicates greater equity.
\end{itemize}

\paragraph{\textbf{Comprehensive Recommendation Quality.}}
In real-world applications, optimizing for fairness alone may degrade standard accuracy metrics, while optimizing for accuracy alone may exacerbate unfairness. For concise comparison, we use $Q$ as an \textbf{comprehensive recommendation quality} that combines recommendation performance, cross-group alignment, and within-group equity:
\vspace{-5pt}
\begin{equation}
\label{eq:q}
Q = \bar{M} \cdot \frac{\mathrm{UGF}}{1 + \mathrm{UIF}},
\end{equation}

where $\bar{M} = \frac{1}{|\mathcal{U}|}\sum_{u_x \in \mathcal{U}} M_{u_x}$ denotes the average performance metric across all users. The multiplicative form is deliberately non-compensatory: weak utility or weak cross-group alignment proportionally limits the summary score, so strength in one dimension cannot fully offset a poor value in another. The term $1+\mathrm{UIF}$ converts the nonnegative within-group disparity into a decreasing factor while preserving its ordering. We do not claim that this scalarization is the unique normative definition of contribution-performance fairness; it is an interpretable summary used together with separately reported $\bar M$, UGF, and UIF.
\vspace{-3pt}
\paragraph{\textbf{Problem Formulation.}}
Given the user set $\mathcal{U}$, the item set $\mathcal{I}$, and the user-item interaction matrix $\mathbf{O}$, CPFR learns model parameters $\Theta$ by jointly optimizing $\mathcal{L}_{\mathrm{rec}}$, $\mathcal{L}_{\mathrm{UIF}}$, and $\mathcal{L}_{\mathrm{UGF}}$ as defined in Sec.~4.

\section{METHODOLOGY}

\subsection{Overview}
We propose the \textbf{Contribution-Performance Fair Recommender (CPFR)}, a unified framework that integrates contribution-performance fairness into standard recommendation training. CPFR consists of two cascaded components:

\vspace{-3pt}
\paragraph{Component I: Contribution Assessment and Group Partitioning.}
A standard base recommender (e.g., NCF~\cite{he2017ncf}) is first trained on $\mathbf{O}_{\text{train}}$. Its per-user training dynamics are used to estimate $C_{u_x}$, after which users are partitioned into $J$ ordered contribution groups for subsequent fairness optimization.

\paragraph{Component II: Multi-Objective Fairness Optimization.}
Using the contribution groups from Component I, the final model jointly optimizes $\mathcal{L}_{\mathrm{rec}}$, $\mathcal{L}_{\mathrm{UIF}}$, and $\mathcal{L}_{\mathrm{UGF}}$.

\vspace{-7pt}
\subsection{Contribution Assessment and Group Partitioning}
This component aims to quantify user contributions and construct statistically meaningful user groups. Firstly, we train a base recommendation model (e.g., NCF) on $\mathbf{O}_{\text{train}}$ by minimizing the standard Binary Cross-Entropy (BCE) loss.
Consistent with findings in~\cite{Paul2021DeepLO}, we collect the three signals defined in Sec.~3.2: interaction volume ($\mathrm{Vol}_{u_x}$), loss alignment ($\mathrm{Align}_{u_x}$), and optimization intensity ($\mathrm{Grad}_{u_x}$) during the early training epochs. These signals are normalized at the population level and combined via Eq.~\eqref{eq:contribution} to obtain the estimated training-contribution score $C_{u_x}$. Then users are partitioned into $J$ disjoint groups $\mathcal{G} = \{G_j\}_{j=1}^{J}$ based on the estimated $C_{u_x}$ values. To ensure statistically coherent groups, we employ the Jenks Natural Breaks Optimization method~\cite{jenks1967data}. This approach minimizes within-group variance of contribution scores while maximizing between-group separation, thereby forming distinct contribution groups for subsequent fairness optimization. Because Jenks determines breakpoints from the empirical contribution distribution rather than fixed quantiles, the resulting groups may contain different numbers of users.

\vspace{-6pt}
\subsection{Multi-Objective Fairness Optimization}
Building upon the user groups established in Component I, we formulate the core training objective of CPFR to unify recommendation accuracy and contribution-based fairness. The total loss comprises three components:
\vspace{-3pt}
\paragraph{\textbf{Recommendation Loss ($\mathcal{L}_{\mathrm{rec}}$).}}
This is the standard recommendation objective. In this work, we adopt the Binary Cross-Entropy (BCE) loss.
\vspace{-3pt}
\paragraph{\textbf{Within-Group Equity Loss ($\mathcal{L}_{\mathrm{UIF}}$).}}
To enforce within-group equity, we minimize within-group BCE-loss dispersion among users in the same contribution group. Let $\bar{\ell}_{u_x}$ denote the average BCE loss of user $u_x$ over their interactions in the current batch. The UIF loss is defined as the pairwise dispersion within each group:
\begin{equation}
\label{eq:uif_loss}
\mathcal{L}_{\mathrm{UIF}} = \sum_{j=1}^{J} \frac{1}{n_j} \sum_{\substack{u_x,u_y \in G_j, x<y}} (\bar{\ell}_{u_x} - \bar{\ell}_{u_y})^2.
\end{equation}
We normalize each group by $n_j$ rather than applying a single global factor $1/N_{\mathrm{pairs}}$ to all within-group pairs, where $N_{\mathrm{pairs}}=\sum_{j=1}^{J}\binom{n_j}{2}$. Under global pair normalization, a group's pair count grows quadratically with $n_j$, causing large groups to dominate the aggregate loss and weakening gradient signals from small groups. Dividing each group by $n_j$ reduces this dependence to approximately linear in group size, preserving effective optimization signals for small groups. Thus, $\mathcal{L}_{\mathrm{UIF}}$ is a differentiable proxy for within-group loss dispersion. In minibatch training, the displayed full-data quantities are estimated from the minibatch.

\paragraph{\textbf{Cross-Group Alignment Loss ($\mathcal{L}_{\mathrm{UGF}}$).}}
To enforce monotonic alignment between contribution and recommendation performance, we penalize violations where a higher-contribution group incurs higher loss (i.e., worse performance) than a lower-contribution one. Let the groups be ordered by their contribution as $C_{G_1} > C_{G_2} > \cdots > C_{G_J}$ and let $\bar{\ell}_{G_j}$ denote the average loss of group $G_j$. We define the UGF loss as:
\begin{equation}
\label{eq:ugf_loss}
\mathcal{L}_{\mathrm{UGF}} = \sum_{j=1}^{J-1} \max\left(0, \; \bar{\ell}_{G_j} - \bar{\ell}_{G_{j+1}} \right),
\end{equation}
This loss encourages the ordering $\bar{\ell}_{G_1} \leq \bar{\ell}_{G_2} \leq \cdots \leq \bar{\ell}_{G_J}$, ensuring that groups contributing more receive no worse performance compared with the ones with less contribution.
\vspace{-3pt}

\paragraph{\textbf{Overall Objective and Optimization.}}
The final training objective is a weighted combination of the three losses:
\begin{equation}
\label{eq:total_loss}
\mathcal{L}_{\mathrm{total}} = \mathcal{L}_{\mathrm{rec}} + \lambda_1 \mathcal{L}_{\mathrm{UIF}} + \lambda_2 \mathcal{L}_{\mathrm{UGF}}.
\end{equation}
Since these objectives may induce conflicting gradients, we employ MoCo~\cite{moco2023}, a provably convergent multi-objective optimization algorithm, to dynamically adjust gradient directions and mitigate interference. This ensures convergence to a Pareto-stationary solution that balances recommendation accuracy, cross-group alignment, and within-group equity without prioritizing any single objective prematurely.

\vspace{-6pt}

\subsection{Theoretical Analysis of Incentives}
\label{sec:theory}
We provide a concise game-theoretic analysis to illustrate the incentive advantage of contribution-aligned allocation. Consider $N\geq2$ homogeneous theoretical users, indexed by $r\in\{1,\dots,N\}$. Each user chooses a data contribution $c_r\geq0$ at cost $\phi(c_r)$, where $\phi$ is differentiable and strictly convex. The total theoretical contribution is $C_{\mathrm{tot}}=\sum_{r=1}^{N}c_r$, and the aggregate recommendation performance is $A(C_{\mathrm{tot}})$, where $A(0)=0$, $A'(C_{\mathrm{tot}})>0$ denotes a positive marginal performance gain, and $A''(C_{\mathrm{tot}})<0$ denotes diminishing returns. Let $p_r$ be the recommendation payoff allocated to user $r$; each user maximizes their net utility:
\[
V_r=p_r-\phi(c_r).
\]
We compare two idealized allocation rules:

\begin{itemize}[noitemsep,topsep=2pt]
    \item \textbf{Equal-Performance (EP):} $p_r^{\mathrm{EP}}=A(C_{\mathrm{tot}})/N$.
    \item \textbf{Contribution-Proportional (CP):} $p_r^{\mathrm{CP}}=(c_r/C_{\mathrm{tot}})A(C_{\mathrm{tot}})$.
\end{itemize}

\paragraph{\textit{Nash Equilibrium} Conditions.}
Assume that both allocation games admit a unique positive symmetric Nash equilibrium. Let $c^*_{\mathrm{EP}}$ and $c^*_{\mathrm{CP}}$ denote the equilibrium per-user contributions. Under EP, a user's marginal benefit from contribution is only a $1/N$ share of the marginal performance gain:
\begin{equation}
\frac{1}{N} A'\!\left(Nc^*_{\mathrm{EP}}\right)
=
\phi'\!\left(c^*_{\mathrm{EP}}\right).
\tag{E1}
\end{equation}

Under CP, differentiating the allocated performance with respect to $c_r$ gives
\[
\frac{\partial V_r^{\mathrm{CP}}}{\partial c_r}
=
\frac{C_{\mathrm{tot}}-c_r}{C_{\mathrm{tot}}^2}A(C_{\mathrm{tot}})
+
\frac{c_r}{C_{\mathrm{tot}}}A'(C_{\mathrm{tot}})
-
\phi'(c_r).
\]
Substituting the symmetric profile $c_r=c^*_{\mathrm{CP}}$ and $C_{\mathrm{tot}}=Nc^*_{\mathrm{CP}}$ yields the following simplified equilibrium condition:
\begin{equation}
\frac{N-1}{N}\cdot
\frac{A\!\left(Nc^*_{\mathrm{CP}}\right)}
{Nc^*_{\mathrm{CP}}}
+
\frac{1}{N} A'\!\left(Nc^*_{\mathrm{CP}}\right)
=
\phi'\!\left(c^*_{\mathrm{CP}}\right).
\tag{E2}
\end{equation}

\paragraph{Comparison.}
For any common positive contribution level $c$, CP provides an additional marginal incentive:
\[
\underbrace{
\left[
\frac{N-1}{N}\frac{A(Nc)}{Nc}
+
\frac{1}{N}A'(Nc)
\right]}_{\text{CP marginal benefit}}
-
\underbrace{
\frac{1}{N}A'(Nc)
}_{\text{EP marginal benefit}}
=
\frac{N-1}{N}\frac{A(Nc)}{Nc}
>0.
\]
Because $A$ is concave and $\phi$ is strictly convex, the marginal-benefit--marginal-cost conditions admit the stated unique crossing. Hence,
\[
c^*_{\mathrm{CP}}>c^*_{\mathrm{EP}},
\qquad
A\!\left(Nc^*_{\mathrm{CP}}\right)
>
A\!\left(Nc^*_{\mathrm{EP}}\right).
\]

\paragraph{Interpretation.}
Under CP, an additional contribution improves both a user's share of system performance and the aggregate recommendation performance itself. This stronger marginal incentive induces higher equilibrium contribution and, consequently, higher aggregate recommendation performance. CPFR does not implement exact proportional allocation; rather, its group-level objective encourages the corresponding contribution--performance ordering. The analysis therefore provides an idealized incentive rationale for contribution-aligned recommendation.

\section{EXPERIMENTS}
We evaluate CPFR through the following questions:

\begin{itemize}
    \item \textbf{RQ1}: How does CPFR compare with state-of-the-art fairness frameworks in $Q$, NDCG, and HR? 

    \item \textbf{RQ2}: How does CPFR make trade-offs among recommendation accuracy, cross-group alignment (UGF), and within-group equity (UIF)? 

    \item \textbf{RQ3}: How do grouping strategies affect recommendation accuracy, UGF, and UIF?

    \item \textbf{RQ4}: How do the hyper-parameters affect the CPFR's ability to maintain high $Q$?
\end{itemize}
\vspace{-8pt}
\subsection{Experimental Setup}
\subsubsection{Datasets}
We evaluate CPFR on three publicly available benchmark datasets: \textbf{MovieLens-1M}~\cite{harper2015movielens}, \textbf{MovieLens-100K}~\cite{harper2015movielens} and \textbf{Amazon Office}~\cite{mcauley2015image}.  As previous
work did~\cite{zhu2024afrl}, we treat interactions with rating $\geq 4$ as positive samples and filter users with fewer than 10 interactions for each dataset. The detailed statistics about datasets are summarized in Table~\ref{tab:datasets}.

\begin{table}[!htbp]
\caption{Statistics of Datasets}
\label{tab:datasets}
\begin{tabular}{lcccc}
\toprule
\textbf{Dataset} & \textbf{Users} & \textbf{Items} & \textbf{Interactions} & \textbf{Density} \\
\midrule
MovieLens-100K & 943 & 1682 & 100000 & 6.305\% \\
MovieLens-1M & 6040 & 3706 & 1000209 & 4.468\% \\
Amazon Office & 101501 & 27965 & 800357 & 0.028\% \\
\bottomrule
\end{tabular}
\end{table}

\vspace{-8pt}
\subsubsection{Baselines}
We compare CPFR with a vanilla model and three state-of-the-art fairness-aware recommenders:

\begin{itemize}
    \item \textbf{Vanilla}: The backbone trained with BCE loss and no fairness constraint.
    
    \item \textbf{AFRL}~\cite{zhu2024afrl}: An adversarial method that removes sensitive-attribute information from user representations.
    
    \item \textbf{CUGF}~\cite{bisht2025cugf}: A confidence-aware method with user-specific fairness weights and variable-length recommendation lists.
    
    \item \textbf{FairNS}~\cite{fairns2025}: A stratified negative-sampling method with group-specific sampling ratios.
\end{itemize}

We implement above methods with three backbones: NCF~\cite{he2017ncf}, DCN~\cite{wang2017dcn}, and DeepFM~\cite{guo2017deepfm}.

\vspace{-3pt}
\subsubsection{Evaluation Protocols}
Following~\cite{zhu2024afrl}, we use leave-one-out evaluation with NDCG@$K$ and HR@$K$. UIF, UGF, and $Q$ which are defined in Section~\ref{subsec:contribution-fairness} are computed from NDCG@$K$; higher $Q$ indicates a better accuracy--fairness balance. All results are averaged over 5 random experimental runs to ensure statistical reliability. 

\vspace{-3pt}
\subsubsection{Implementation Details}
\label{subsec:implementation_details}

Due to the distinct characteristics of fairness frameworks, it is necessary to adopt differentiated evaluation protocols to ensure fair and meaningful comparisons.

\vspace{-5pt}
\paragraph{Unified Settings.}
For validation and testing, we sample one positive and 99 negatives per user; remaining interactions are used for training. We use batch size 2048, learning rate $1\times10^{-3}$, embedding dimension 64, Adam as the optimizer, and early stopping on validation NDCG@10 with patience 10.
\vspace{-5pt}
\paragraph{Framework-Specific Settings.}
To ensure fair comparison against different types of fairness-aware methods, we adopt tailored evaluation settings for each baseline:

\begin{itemize}
    \item \textbf{CPFR vs. Vanilla \& AFRL}: We use one training negative per positive, 99 validation/test negatives, and $K=10$, following AFRL ~\cite{zhu2024afrl}.

    \item \textbf{CPFR vs. FairNS}: FairNS uses group-specific ratios of 1--8 negatives per positive. For matched training scale, CPFR uses $n_{\mathrm{neg}}=\lfloor r\rfloor$ negatives per positive, where $r$ is FairNS's average number of sampled negatives per positive training interaction; $K=10$.
    
    \item \textbf{CPFR vs. CUGF}: 
    Because CUGF produces variable-length lists for different users, both methods are evaluated at $K=\lfloor L\rfloor$, where $L$ is CUGF's mean list length.
\end{itemize}

For CPFR, we use a larger number of user groups $J$ for datasets with more users. The $J$ is set to 10 for MovieLens-100K, 15 for Amazon Office, and 20 for MovieLens-1M. For each dataset--backbone pair, we initialized the hyper-parameter search ranges of AFRL, FairNS, and CUGF using the configurations reported in their original papers, and then re-tuned their method-specific hyper-parameters on the validation set. CPFR's $\lambda_1$ and $\lambda_2$ were tuned under the same validation criterion. For every method, we selected the configuration with the highest validation $Q$, and evaluated the selected configuration on the test set. Thus, $Q$ is used consistently as the operational model selection criterion for all compared methods, rather than only favoring CPFR.

\vspace{-8pt}
\subsection{Main Results (RQ1)}
\label{sec:main_results}
Tables~\ref{tab:main_results_movielens100k}, ~\ref{tab:main_results_movielens1m} and ~\ref{tab:main_results_amaoffice} present the performance of CPFR and baseline fair models on the Movielens-100K, Movielens-1M, and Office datasets, respectively. Regarding the comprehensive recommendation quality $Q$, CPFR achieves the best trade-off in almost all cases. Quantitatively, CPFR outperforms AFRL in 8 out of 9 configurations, maintaining over 90\% of vanilla accuracy while AFRL collapses to merely $\sim$40\%. Against FairNS, CPFR keeps the accuracy gap within 3\% while securing superior User Group Fairness ($\text{UGF}$) in most cases. Compared to CUGF, CPFR attains strictly higher $Q$ values across all settings, demonstrating comprehensive robustness.

\begin{table*}[!htbp]
\centering
\caption{Performance comparison on Movielens100k. $\text{Vanilla}$ denotes backbone model trained without fairness constraints. $n_{\mathrm{neg}}$ indicates the training negative sampling ratio, and $K$ indicates the evaluation list length. $\downarrow$ indicates lower is better, $\uparrow$ indicates higher is better.}
\label{tab:main_results_movielens100k}
\scriptsize
\begin{threeparttable}
\resizebox{0.75\textwidth}{!}{
\begin{tabular}{ll|cc|ccccc}
\specialrule{1.2pt}{0pt}{0pt}
\textbf{Backbone} & \textbf{Method} & \textbf{$n_{\mathrm{neg}}$} & \textbf{$K$} & NDCG@K$\uparrow$ & HR@K$\uparrow$ & UIF@K$\downarrow$ & UGF@K$\uparrow$ & Q@K$\uparrow$ \\
\specialrule{1.2pt}{0pt}{0pt}
\multirow{7}{*}{NCF}
& Vanilla & 1 & 10 & \textbf{0.3633±0.0120} & \textbf{0.6192±0.0181} & 0.2435±0.0075 & 0.4857±0.0764 & 0.1417±0.0211 \\
& AFRL & 1 & 10 & 0.1513±0.0148 & 0.2934±0.0233 & \textbf{0.1428±0.0160} & 0.3571±0.2396 & 0.0471±0.0317 \\
& CPFR & 1 & 10 & 0.3463±0.0151 & 0.5978±0.0213 & 0.2401±0.0091 & \textbf{0.5595±0.1268} & \textbf{0.1561±0.0359} \\
\cmidrule(lr){2-9}
& FairNS & 1-8 & 10 & \textbf{0.3994±0.0135} & \textbf{0.6682±0.0238} & 0.2457±0.0100 & 0.4000±0.0581 & 0.1287±0.0223 \\
& CPFR & 3 & 10 & 0.3886±0.0079 & 0.6589±0.0246 & \textbf{0.2402±0.0131} & \textbf{0.4167±0.0679} & \textbf{0.1305±0.0210} \\
\cmidrule(lr){2-9}
& CUGF & 1 & 3 & \textbf{0.2826±0.0278} & 0.3844±0.0555 & 0.2871±0.0084 & 0.4857±0.0987 & 0.1060±0.0210 \\
& CPFR & 1 & 3 & 0.2789±0.0248 & \textbf{0.3960±0.0494} & \textbf{0.2757±0.0113} & \textbf{0.5476±0.0978} & \textbf{0.1188±0.0169} \\
\midrule
\addlinespace[0.2ex]
\midrule
\multirow{7}{*}{DCN}
& Vanilla & 1 & 10 & \textbf{0.3344±0.0064} & \textbf{0.6016±0.0163} & 0.2235±0.0045 & 0.4286±0.2071 & 0.1171±0.0566 \\
& AFRL & 1 & 10 & 0.2576±0.0460 & 0.4725±0.0819 & \textbf{0.1990±0.0304} & 0.2714±0.2822 & 0.0637±0.0717 \\
& CPFR & 1 & 10 & 0.3191±0.0250 & 0.5918±0.0384 & 0.2110±0.0123 & \textbf{0.4929±0.2356} & \textbf{0.1310±0.0882} \\
\cmidrule(lr){2-9}
& FairNS & 1-8 & 10 & 0.3848±0.0047 & \textbf{0.6611±0.0085} & \textbf{0.2400±0.0109} & 0.2071±0.0933 & 0.0643±0.0289 \\
& CPFR & 3 & 10 & \textbf{0.4171±0.0914} & 0.6457±0.0330 & 0.2795±0.0782 & \textbf{0.3464±0.1191} & \textbf{0.1231±0.0356} \\
\cmidrule(lr){2-9}
& CUGF & 1 & 4 & \textbf{0.2761±0.0159} & 0.4223±0.0466 & 0.2551±0.0091 & 0.3607±0.1896 & 0.0795±0.0430 \\
& CPFR & 1 & 4 & 0.2682±0.0319 & \textbf{0.4312±0.0625} & \textbf{0.2375±0.0159} & \textbf{0.5143±0.1561} & \textbf{0.1131±0.0597} \\
\midrule
\addlinespace[0.2ex]
\midrule
\multirow{7}{*}{DeepFM}
& Vanilla & 1 & 10 & \textbf{0.3355±0.0233} & 0.5931±0.0238 & 0.2322±0.0186 & 0.3429±0.1832 & 0.0927±0.0485 \\
& AFRL & 1 & 10 & 0.3232±0.0305 & 0.5891±0.0299 & \textbf{0.2173±0.0214} & 0.4071±0.2148 & 0.1061±0.0526 \\
& CPFR & 1 & 10 & 0.3303±0.0056 & \textbf{0.5967±0.0070} & 0.2254±0.0019 & \textbf{0.5804±0.1402} & \textbf{0.1567±0.0393} \\
\cmidrule(lr){2-9}
& FairNS & 1-8 & 10 & \textbf{0.3878±0.0084} & \textbf{0.6551±0.0164} & 0.2498±0.0037 & 0.2429±0.0774 & 0.0757±0.0256 \\
& CPFR & 3 & 10 & 0.3787±0.0042 & 0.6421±0.0042 & \textbf{0.2483±0.0062} & \textbf{0.2607±0.0732} & \textbf{0.0791±0.0433} \\
\cmidrule(lr){2-9}
& CUGF & 1 & 5 & 0.2828±0.0273 & 0.4375±0.0425 & 0.2562±0.0254 & 0.3750±0.1567 & 0.0836±0.0316 \\
& CPFR & 1 & 5 & \textbf{0.2854±0.0117} & \textbf{0.4529±0.0332} & \textbf{0.2522±0.0062} & \textbf{0.6339±0.2358} & \textbf{0.1462±0.0593} \\
\specialrule{1.2pt}{0pt}{0pt}
\end{tabular}
}
\end{threeparttable}
\end{table*}
 
\begin{table*}[!htbp]
\centering
\caption{Performance comparison on Movielens1M.}
\label{tab:main_results_movielens1m}
\scriptsize
\begin{threeparttable}
\resizebox{0.75\textwidth}{!}{
\begin{tabular}{ll|cc|ccccc}
\specialrule{1.2pt}{0pt}{0pt}
\textbf{Backbone} & \textbf{Method} & \textbf{$n_{\mathrm{neg}}$} & \textbf{$K$} & NDCG@K$\uparrow$ & HR@K$\uparrow$ & UIF@K$\downarrow$ & UGF@K$\uparrow$ & Q@K$\uparrow$ \\
\specialrule{1.2pt}{0pt}{0pt}
\multirow{7}{*}{NCF}
& Vanilla & 1 & 10 & \textbf{0.3949±0.0148} & 0.6589±0.0169 & 0.2477±0.0073 & 0.6307±0.0450 & 0.1996±0.0154 \\
& AFRL & 1 & 10 & 0.1376±0.0696 & 0.2744±0.1225 & \textbf{0.1287±0.0454} & 0.6125±0.1588 & 0.0784±0.0513 \\
& CPFR & 1 & 10 & 0.3932±0.0027 & \textbf{0.6628±0.0087} & 0.2447±0.0063 & \textbf{0.6682±0.0466} & \textbf{0.2111±0.0151} \\
\cmidrule(lr){2-9}
& FairNS & 1-8 & 10 & \textbf{0.4484±0.0048} & \textbf{0.7210±0.0070} & \textbf{0.2529±0.0043} & \textbf{0.6207±0.0300} & \textbf{0.2222±0.0119} \\
& CPFR & 4 & 10 & 0.4454±0.0054 & 0.7174±0.0103 & 0.2589±0.0041 & 0.6039±0.0429 & 0.2137±0.0152 \\
\cmidrule(lr){2-9}
& CUGF & 1 & 4 & 0.3193±0.0162 & 0.4351±0.0193 & 0.2988±0.0099 & 0.6357±0.0398 & 0.1564±0.0147 \\
& CPFR & 1 & 4 & \textbf{0.3194±0.0102} & \textbf{0.4432±0.0265} & \textbf{0.2949±0.0074} & \textbf{0.6686±0.0536} & \textbf{0.1651±0.0167} \\
\midrule
\addlinespace[0.2ex]
\midrule
\multirow{7}{*}{DCN}
& Vanilla & 1 & 10 & \textbf{0.4109±0.0846} & \textbf{0.6249±0.0182} & 0.2876±0.0907 & 0.6536±0.3305 & 0.1969±0.0966 \\
& AFRL & 1 & 10 & 0.1729±0.0983 & 0.3248±0.1774 & \textbf{0.1509±0.0550} & 0.3696±0.2910 & 0.0719±0.0778 \\
& CPFR & 1 & 10 & 0.3490±0.0133 & 0.6082±0.0246 & 0.2281±0.0061 & \textbf{0.7039±0.1227} & \textbf{0.2008±0.0416} \\
\cmidrule(lr){2-9}
& FairNS & 1-8 & 10 & 0.4245±0.0116 & 0.6955±0.0125 & \textbf{0.2605±0.0050} & \textbf{0.6886±0.1069} & 0.2313±0.0314 \\
& CPFR & 4 & 10 & \textbf{0.5574±0.1528} & \textbf{0.7467±0.0413} & 0.2944±0.0465 & 0.6111±0.1703 & \textbf{0.2495±0.0410} \\
\cmidrule(lr){2-9}
& CUGF & 1 & 4 & \textbf{0.3790±0.1373} & \textbf{0.5411±0.1818} & 0.2963±0.0260 & 0.6514±0.2913 & 0.1670±0.0592 \\
& CPFR & 1 & 4 & 0.3203±0.0566 & 0.5325±0.2049 & \textbf{0.2392±0.0473} & \textbf{0.6939±0.1112} & \textbf{0.1810±0.0499} \\
\midrule
\addlinespace[0.2ex]
\midrule
\multirow{7}{*}{DeepFM}
& Vanilla & 1 & 10 & \textbf{0.3787±0.0075} & 0.6445±0.0158 & 0.2441±0.0046 & 0.7361±0.0543 & 0.2241±0.0178 \\
& AFRL & 1 & 10 & 0.2532±0.0608 & 0.4430±0.0839 & \textbf{0.2098±0.0409} & 0.4493±0.1814 & 0.0924±0.0365 \\
& CPFR & 1 & 10 & 0.3751±0.0119 & \textbf{0.6487±0.0099} & 0.2342±0.0106 & \textbf{0.7368±0.0798} & \textbf{0.2243±0.0292} \\
\cmidrule(lr){2-9}
& FairNS & 1-8 & 10 & \textbf{0.4413±0.0044} & \textbf{0.7094±0.0070} & 0.2692±0.0031 & \textbf{0.6871±0.0636} & \textbf{0.2390±0.0231} \\
& CPFR & 4 & 10 & 0.4082±0.0516 & 0.6694±0.0656 & \textbf{0.2552±0.0077} & 0.6321±0.1133 & 0.2020±0.0133 \\
\cmidrule(lr){2-9}
& CUGF & 1 & 4 & 0.3098±0.0082 & 0.4358±0.0240 & 0.2897±0.0085 & \textbf{0.7246±0.0565} & 0.1741±0.0147 \\
& CPFR & 1 & 4 & \textbf{0.3160±0.0124} & \textbf{0.4664±0.0118} & \textbf{0.2743±0.0132} & 0.7193±0.0857 & \textbf{0.1786±0.0240} \\
\specialrule{1.2pt}{0pt}{0pt}
\end{tabular}
}
\end{threeparttable}
\end{table*}
 
\begin{table*}[!htbp]
\centering
\caption{Performance comparison on Amazon Office.}
\label{tab:main_results_amaoffice}
\scriptsize
\begin{threeparttable}
\resizebox{0.75\textwidth}{!}{
\begin{tabular}{ll|cc|ccccc}
\specialrule{1.2pt}{0pt}{0pt}
\textbf{Backbone} & \textbf{Method} & \textbf{$n_{\mathrm{neg}}$} & \textbf{$K$} & NDCG@K$\uparrow$ & HR@K$\uparrow$ & UIF@K$\downarrow$ & UGF@K$\uparrow$ & Q@K$\uparrow$ \\
\specialrule{1.2pt}{0pt}{0pt}
\multirow{7}{*}{NCF}
& Vanilla & 1 & 10 & \textbf{0.2299±0.0437} & \textbf{0.3919±0.0630} & 0.2337±0.0247 & 0.5536±0.1195 & 0.1034±0.0297 \\
& AFRL & 1 & 10 & 0.0704±0.0167 & 0.1416±0.0232 & \textbf{0.0968±0.0415} & 0.5573±0.1881 & 0.0358±0.0132 \\
& CPFR & 1 & 10 & 0.2131±0.0426 & 0.3592±0.0586 & 0.2226±0.0249 & \textbf{0.6236±0.1905} & \textbf{0.1088±0.0387} \\
\cmidrule(lr){2-9}
& FairNS & 1-8 & 10 & \textbf{0.2307±0.0294} & \textbf{0.3896±0.0297} & 0.2478±0.0211 & 0.4009±0.1200 & 0.0730±0.0193 \\
& CPFR & 7 & 10 & 0.1040±0.0133 & 0.1903±0.0224 & \textbf{0.1371±0.0245} & \textbf{0.8545±0.1000} & \textbf{0.0781±0.0126} \\
\cmidrule(lr){2-9}
& CUGF & 1 & 5 & \textbf{0.1963±0.0486} & 0.2889±0.0757 & 0.2424±0.0281 & 0.5318±0.0990 & 0.0846±0.0293 \\
& CPFR & 1 & 5 & 0.1917±0.0476 & \textbf{0.2914±0.0719} & \textbf{0.2270±0.0265} & \textbf{0.6382±0.1456} & \textbf{0.0995±0.0316} \\
\midrule
\addlinespace[0.2ex]
\midrule
\multirow{7}{*}{DCN}
& Vanilla & 1 & 10 & \textbf{0.2466±0.0475} & \textbf{0.3975±0.0552} & 0.2539±0.0275 & 0.3982±0.1444 & 0.0786±0.0294 \\
& AFRL & 1 & 10 & 0.1742±0.0361 & 0.3021±0.0466 & \textbf{0.2256±0.0418} & \textbf{0.4364±0.0937} & 0.0621±0.0178 \\
& CPFR & 1 & 10 & 0.2411±0.0468 & 0.3900±0.0565 & 0.2505±0.0284 & 0.4209±0.1310 & \textbf{0.0813±0.0275} \\
\cmidrule(lr){2-9}
& FairNS & 1-8 & 10 & 0.2534±0.0509 & 0.4072±0.0593 & 0.2543±0.0304 & 0.3455±0.1306 & 0.0694±0.0290 \\
& CPFR & 7 & 10 & \textbf{0.2582±0.0486} & \textbf{0.4207±0.0570} & \textbf{0.2538±0.0304} & \textbf{0.4900±0.0314} & \textbf{0.1007±0.0188} \\
\cmidrule(lr){2-9}
& CUGF & 1 & 6 & 0.2199±0.0518 & 0.3129±0.0681 & 0.2631±0.0307 & 0.3727±0.0817 & 0.0640±0.0146 \\
& CPFR & 1 & 6 & \textbf{0.2221±0.0500} & \textbf{0.3281±0.0665} & \textbf{0.2577±0.0311} & \textbf{0.4009±0.1442} & \textbf{0.0710±0.0279} \\
\midrule
\addlinespace[0.2ex]
\midrule
\multirow{7}{*}{DeepFM}
& Vanilla & 1 & 10 & \textbf{0.2465±0.0477} & \textbf{0.3984±0.0573} & 0.2505±0.0335 & 0.4345±0.1278 & 0.0848±0.0235 \\
& AFRL & 1 & 10 & 0.1626±0.0202 & 0.2656±0.0469 & \textbf{0.2196±0.0856} & 0.4455±0.2186 & 0.0592±0.0301 \\
& CPFR & 1 & 10 & 0.2417±0.0465 & 0.3943±0.0570 & 0.2472±0.0272 & \textbf{0.4636±0.0888} & \textbf{0.0896±0.0219} \\
\cmidrule(lr){2-9}
& FairNS & 1-8 & 10 & 0.2596±0.0507 & 0.4145±0.0597 & 0.2622±0.0300 & 0.3918±0.1381 & 0.0803±0.0330 \\
& CPFR & 7 & 10 & \textbf{0.2608±0.0517} & \textbf{0.4211±0.0606} & \textbf{0.2564±0.0322} & \textbf{0.4882±0.0988} & \textbf{0.0990±0.0158} \\
\cmidrule(lr){2-9}
& CUGF & 1 & 6 & \textbf{0.2233±0.0476} & 0.3239±0.0567 & 0.2589±0.0366 & 0.4636±0.1285 & 0.0807±0.0204 \\
& CPFR & 1 & 6 & 0.2204±0.0507 & \textbf{0.3253±0.0695} & \textbf{0.2541±0.0286} & \textbf{0.4718±0.0987} & \textbf{0.0812±0.0150} \\
\specialrule{1.2pt}{0pt}{0pt}
\end{tabular}
}
\end{threeparttable}
\end{table*}

As to the dimensions of $Q$, AFRL achieves the lowest UIF scores, indicating the strongest within-group equity, yet this comes at the cost of drastic accuracy degradation. In contrast, CPFR preserves discriminative power while improving cross-group alignment, significantly boosting $\text{UGF}$ without compromising $\text{UIF}$. While FairNS maintains competitive accuracy, it exhibits limited $\text{UGF}$ improvement; CPFR surpasses it in $\text{UGF}$ while maintaining accuracy parity. Furthermore, CPFR consistently dominates CUGF in both $\text{UIF}$ and $\text{UGF}$ simultaneously. Thus, CPFR minimizes accuracy loss while jointly improving within-group equity and cross-group alignment.

\vspace{-9pt}
\subsection{Trade-off Analysis (RQ2)}
We analyze the trade-off between fairness and accuracy by tuning $\lambda_1$ and $\lambda_2$ in Eq.~\ref{eq:total_loss}. 
Experiments on Movielens100k with DeepFM span 100 configurations ($\lambda_1, \lambda_2 \in \{0, 0.01, 0.05, 0.1, 0.5, 1.0, 2.0, 3.0, 4.0, 5.0\}$).

\begin{figure*}[!htbp]
    \centering
    \includegraphics[width=\textwidth]{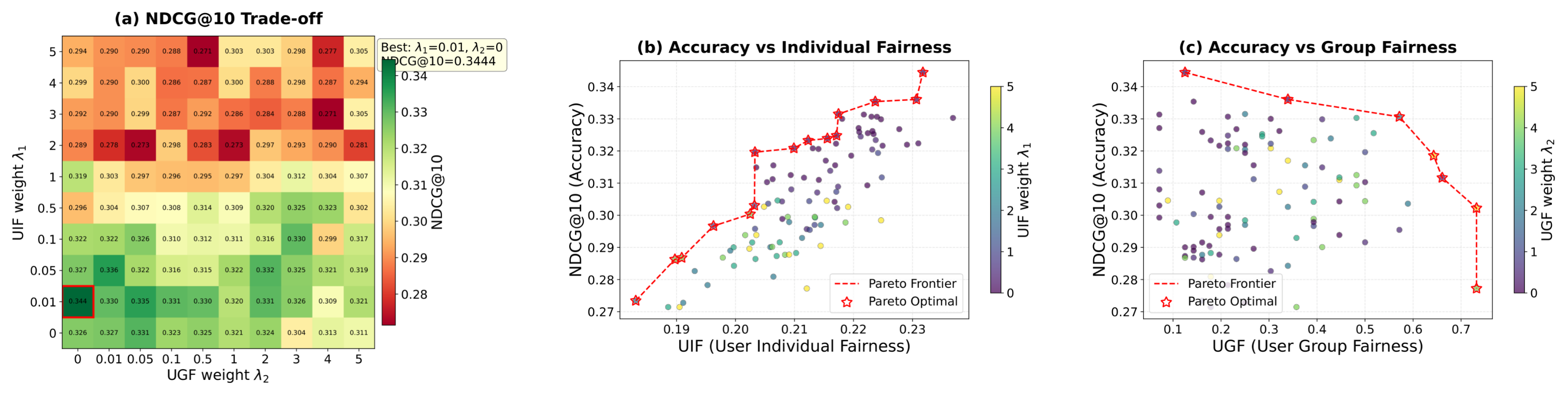}
    \caption{Trade-off analysis between accuracy and fairness under different $\lambda_1$ and $\lambda_2$ configurations. 
    (a) NDCG heatmap showing accuracy variations; 
    (b)(c) Pareto frontier between accuracy and UIF (within-group equity; lower is better) / UGF (cross-group alignment). 
    Red boxes in (a) indicate optimal accuracy configurations ($\lambda_1 = 0.01$, $\lambda_2 = 0$). 
    Red stars in (b)(c) mark Pareto-optimal trade-off points.}
    \label{fig:tradeoff_analysis}
\end{figure*}

Figure~\ref{fig:tradeoff_analysis}(a) shows the NDCG heatmap. Accuracy peaks at minimal fairness weights ($\lambda_1=0.01, \lambda_2=0$, NDCG=$0.3444$), close to the vanilla baseline. Increasing weights systematically degrades accuracy, with high fairness settings reducing NDCG by approximately 5-10\%. Panels (b) and (c) depict Pareto frontiers. In (b), higher $\lambda_1$ consistently reduces UIF, thereby improving within-group equity at the cost of NDCG. Similarly, in (c), increasing $\lambda_2$ improves cross-group alignment (UGF) at the expense of accuracy. The frontiers confirm that neither fairness dimension can be improved without sacrificing accuracy.

These results confirm that $\lambda_1$ and $\lambda_2$ independently control within-group equity and cross-group alignment, respectively. CPFR allows flexible navigation of the accuracy-fairness trade-off via these two knobs, enabling practitioners to select optimal configurations based on domain priorities.

\vspace{-10pt}
\subsection{Ablation Study (RQ3)}
\label{sec:ablation}

This subsection investigates how different configurations of user grouping strategies affect the trade-off among the dimensions of $Q$.

\begin{figure*}[!htbp]
    \centering
    \includegraphics[width=0.9\textwidth]{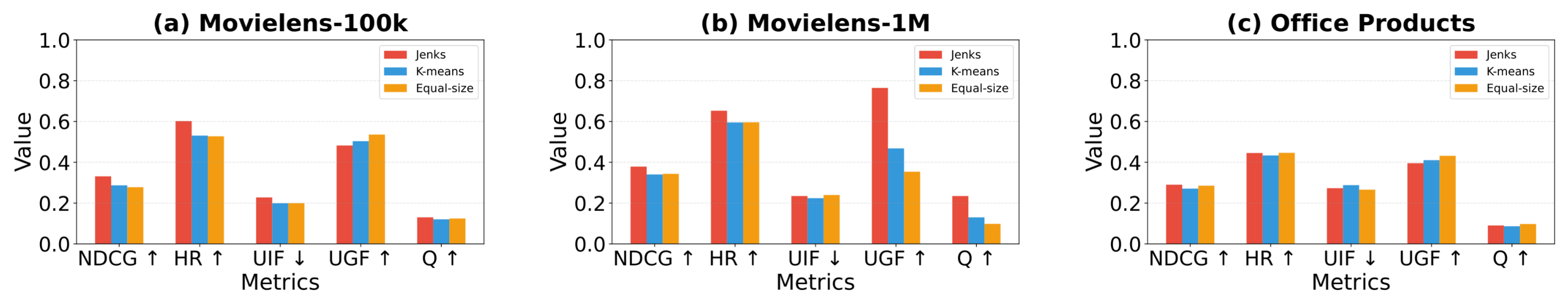}
    \caption{Grouping strategy comparison across three datasets.
    Each subplot shows five metrics: NDCG , HR, UIF, UGF, and Q.
    Arrows indicate optimization direction: $\uparrow$ for metrics to maximize, $\downarrow$ for metrics to minimize.}
    \label{fig:grouping_comparison}
\end{figure*}

For user grouping strategies, we evaluate three grouping methods: Jenks Natural Breaks and K-means clustering  (both adaptively determining group sizes), and Equal-size Partitioning (fixed-size intervals along the contribution axis). As shown in Figure~\ref{fig:grouping_comparison}, distinct trade-offs emerge: Jenks consistently guarantees optimal performance and achieves competitive $Q$ values; K-means suffers from instability due to its stochastic nature; Equal-size maintains balanced performance but often sacrifices accuracy. We select Jenks Natural Breaks prioritizing composite performance ($Q$). While theoretically equivalent to K-means on 1D data, Jenks eliminates stochasticity via deterministic breakpoint detection, offering stable convergence critical for reproducible experiments and precise differentiation of heterogeneous user contributions.

\vspace{-10pt}
\subsection{Hyper-parameter Analysis (RQ4)}
\label{sec:hyperparams}
\begin{figure*}[!htbp]
    \centering
    \includegraphics[width=0.9\textwidth]{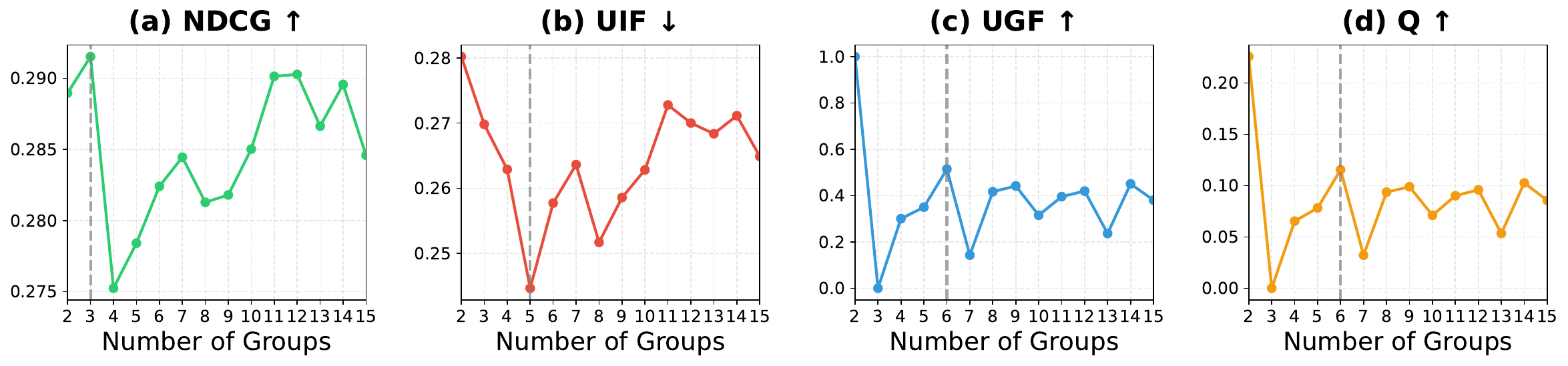}
    \caption{Impact of group number on performance metrics.
    Each subplot shows the variation of a single metric across actual group counts from 2 to 15.
    Gray dashed lines mark the optimal positions.
    The cases of 2 groups (UGF=1.0) is trivial values included for reference but excluded from optimal analysis.}
    \label{fig:groupnumber_ablation}
\end{figure*}

The number of contribution groups $J$ determines the grouping granularity. A smaller $J$ leads to coarser grouping with more users per group, while a larger $J$ results in finer-grained groups with fewer users. We vary $J$ from 2 to 21, resulting in actual group counts ranging from 2 to 15 after merging small groups. Experiments are conducted on the Office dataset with DeepFM as the backbone recommender. Figure~\ref{fig:groupnumber_ablation} illustrates the impact of group granularity $J$ on NDCG, UIF, UGF, and $Q$, revealing a distinct non-linear optimization trajectory across these dimensions.

At low $J$, the model approximates a single-objective optimization regime, where performance loss dominates the objective. This yields high accuracy but poor structural fairness and high individual disparity, resulting in suboptimal $Q$. As $J$ increases moderately, the grouping structure effectively captures nuanced unfairness signals, enabling targeted fairness optimization. This triggers a classic trade-off: NDCG decreases slightly, while both UGF and UIF improve significantly as the model learns to distinguish contribution disparities. However, at excessive $J$, the benefits diminish. Overly fine-grained groups contain too few users, introducing statistical noise and instability. Consequently, fairness optimization becomes difficult, leading to a slight recovery in NDCG (due to reduced fairness constraints) but a marginal decline in overall fairness, ultimately lowering $Q$. Thus, the optimal $J$ balances signal richness against statistical stability, avoiding the pitfalls of both under-fitting (too few groups) and over-fitting/noise (too many groups).

\vspace{-10pt}
\section{Conclusion}
\label{sec:conclusion}

In this paper, we propose a novel framework, called Contribution-Performance Fair Recommender (CPFR), for addressing contribution based unfairness in recommender systems. CPFR explicitly models estimated, training-dependent user contribution and achieves contribution-performance alignment, balancing recommendation accuracy with cross-group alignment and within-group equity. Specifically, CPFR introduces a multi-dimensional quantification mechanism integrating interaction volume, loss alignment, and optimization intensity. By employing Jenks natural breaks, it adaptively stratifies users into ordered groups. With the novel Dual-Fairness Optimization via UGF and UIF losses, CPFR encourages that higher-contribution groups receive no worse performance while minimizing within-group disparity, retaining approximately 90\% of the vanilla model accuracy. Finally, extensive experiments on three benchmark datasets demonstrate the superiority of CPFR. Separately, under an idealized allocation model, our game-theoretic analysis shows that contribution-proportional allocation induces higher equilibrium contribution and aggregate recommendation performance than equal-performance allocation.

\clearpage
\bibliographystyle{ACM-Reference-Format}
\bibliography{CPFR_references}

\end{document}